\documentclass[a4paper,11pt]{article}
\pdfoutput=1

\usepackage{jcappub}
\usepackage[T1]{fontenc}
\usepackage{multirow}
\usepackage{array, makecell}

\title{\boldmath Effects of baryons on weak lensing peak statistics}

\author[1,2]{Andreas J. Weiss,}
\author[1]{Aurel Schneider,}
\author[1]{Raphael Sgier,}
\author[1]{Tomasz Kacprzak,}
\author[1]{Adam Amara,}
\author[1]{and Alexandre Refregier}
\affiliation[1]{Institute for Particle Physics and Astrophysics, ETH Zurich, Wolfgang Pauli Strasse 27, 8093 Zurich, Switzerland}
\affiliation[2]{Space Research Institute, Austrian Academy of Sciences, Schmiedlstra{\ss}e 6, 8042 Graz, Austria}

\emailAdd{andreas.weiss@oeaw.ac.at}
\emailAdd{aurel.schneider@phys.ethz.ch}

\abstract{Upcoming weak-lensing surveys have the potential to become leading cosmological probes provided all systematic effects are under control. Recently, the ejection of gas due to feedback energy from active galactic nuclei (AGN) has been identified as major source of uncertainty, challenging the success of future weak-lensing probes in terms of cosmology. In this paper we investigate the effects of baryons on the number of weak-lensing peaks in the convergence field. Our analysis is based on full-sky convergence maps constructed via light-cones from $N$-body simulations, and we rely on the \emph{baryonic correction model} of \citet{schneider_2018} to model the baryonic effects on the density field.  As a result we find that the baryonic effects strongly depend on the Gaussian smoothing applied to the convergence map. For a DES-like survey setup, a smoothing of $\theta_k\gtrsim8$ arcmin is sufficient to keep the baryon signal below the expected statistical error. Smaller smoothing scales lead to a significant suppression of high peaks (with signal-to-noise above 2), while lower peaks are not affected. The situation is more severe for a Euclid-like setup, where a smoothing of $\theta_k\gtrsim16$ arcmin is required to keep the baryonic suppression signal below the statistical error. Smaller smoothing scales require a full modelling of baryonic effects since both low and high peaks are strongly affected by baryonic feedback.}

\begin{document}
\maketitle

\section{Introduction}

In the currently favoured cosmological model, $\Lambda$CDM, the energy density of the universe is dominated by a cosmological constant ($\Lambda$) and a cold dark matter (CDM) component. The most accurate constraints for this model are obtained from the cosmic microwave background signal which are measured using surveys such as Planck \citep{planck_collaboration_2013, planck_collaboration_2015}. 

Gravitational lensing -- the deflection of light around clustered matter -- offers an additional avenue to test the standard model of cosmology. A particularly promising probe is weak lensing, where slight distortions of galaxy shapes are used to determine the underlying matter density distribution including the invisible dark matter component. Due to the weak nature of this effect, large area surveys observing many millions to billions of galaxies are required, such as KiDS\footnote{\texttt{http://kids.strw.leidenuniv.nl/}}, DES\footnote{\texttt{https://www.darkenergysurvey.org/}}, HSC\footnote{\texttt{https://hsc.mtk.nao.ac.jp/ssp/}} or the planned Euclid\footnote{\texttt{https://www.euclid-ec.org/}}, LSST\footnote{\texttt{https://www.lsst.org/}}, and WFIRST\footnote{\texttt{https://wfirst.gsfc.nasa.gov/}} surveys \citep{kids_survey,des_survey,hsc_survey,euclid_survey,lsst_survey}.

In order to use weak-lensing surveys as cosmological probes, accurate predictions of the density distribution in the universe at both linear and nonlinear scales are required. While cosmological perturbation theories only describe the linear regime, full cosmological $N$-body simulations \citep{skillman_2014,heitmann_2016,schneider_2016,heitmann_2019} or semi-analytical methods including the halo model \citep{peacock_1996, smith_2003, takahashi_2012, mead_2015} are able to describe the nonlinear behaviour of structure formation. However, these approaches only consider gravitational effects describing structure formation as a fully collisionless process.

Recent work based on hydrodynamical simulations has produced mounting evidence that a gravity-only approach is no longer sufficiently accurate to predict the weak-lensing signal. For example, Refs.~\citep[][]{vandaalen_2011, hellwing_2016, mummery_2017, castro_2017, springel_2018, chisari_2018,barreira_2019} show that baryonic effects, specifically the energy input from active galactic nuclei (AGN), have an impact on the matter distribution at cosmologically relevant scales. Note, however, that baryonic feedback effects are not predicted from first principles in hydrodynamical simulations, but they are included as sub-grid effects using semi-analytical recipes. As a consequence, there is no quantitative agreement between different simulations in terms of clustering statistics. Differences in the implementation of sub-grid feedback effects lead to a spread of up to 30 percent in the predictions of the power spectrum for wave-modes between $k\sim 0.1$ and 10 h/Mpc \citep[e.g. Ref.][]{huang_2018,chisari_2019}.

An alternative approach to full hydrodynamical simulations are models that parametrise the effects of baryons on the gravitational clustering. These parametrisations are usually performed at the level of dark matter haloes and the different model parameters are constrained using simulations \citep{rudd_2008, zentner_2008, semboloni_2011, fedeli_2014, mead_2015} or direct observations \citep{schneider_2015,schneider_2018}. The advantage of a model parametrisation is that the baryonic effects can be self-consistently included in methods of cosmological parameter inference.

The present paper relies on the \emph{baryonic correction} (BC) model proposed in Refs. \citep{schneider_2015, schneider_2018}. Gravity-only $N$-body simulation outputs are directly perturbed by slightly displacing particles around halo centres according to a parametrised halo profile which includes a stellar, gas, and dark matter component. Compared to the halo model approach, the BC method is based on a simulated density map, accurately reproducing all the complexity of large-scale structures including triaxial haloes, voids, filaments, etc. As a result, the model is not restricted to 2-point statistics but allows to study all summary statistics describing the matter distribution in the universe.

During the last decade, weak lensing peak statistics have emerged as a simple but powerful complementary tool to extract cosmological information that is not accessible with the angular power spectrum alone \citep{jain_2000,wang_2009,kratochvil_2010,dietrich_2010,liu_2015,
liu_pan_2015,kacprzak_2016,fluri_2018,martinet_2018,shan_2018,davies_2019}. The complementary nature of peak counts is highlighted by the fact that a combined analysis including the angular power spectrum results in considerably tighter cosmological constraints than any of these statistics used on their own \citep{dietrich_2010}. The origin of weak lensing peaks has been closely investigated in Refs.  \citep{yang_2011,liu_2016}. From there we know that the highest peaks are typically caused by one single massive halo. Medium and lower peaks, on the other hand, are generated either by shape noise or by line-of-sight projections of multiple smaller haloes.

In this paper we provide the first detailed study of the AGN driven baryonic effects on the peaks of weak-lensing convergence maps. While former work by Refs.~\citep{yang_2013,osato_2015} also looked at the baryonic effects on weak-lensing peaks, they either did not include AGN feedback effects at all \citep{yang_2013} or relied on a very mild AGN feedback recipe \citep{osato_2015}, potentially underestimating the true baryonic depletion effects. The latter statement is based on the angular power spectra published in Refs.~\citep{yang_2013,osato_2015} showing considerably weaker baryonic effects compared to recent hydrodynamical simulations \citep[e.g.][]{vandaalen_2011, mummery_2017}.

In the present study, we construct the convergence maps from a light-cone based on outputs of $N$-body simulations (which have previously been \emph{baryonified} using the baryonic correction model) assuming an extended redshift distribution for the source galaxies. Together with the baryonic effects, we investigate the role of the galaxy number density and smoothing scale on the weak-lensing peak statistics. We thereby consider two survey configurations based on DES and Euclid. This will give us information on how baryons influence both current and next-generation weak lensing surveys.

The paper is structured as follows: In Sec.~\ref{sec:numerical_methods} we summarise the method to generate convergence maps from \textit{N}-body simulations and we discuss the weak-lensing peak counts based on gravity-only simulations without baryonic effects. Sec.~\ref{sec:baryonic_correction_model} provides a brief introduction to the BC model including three benchmark models that cover realistic scenarios in terms of the expected baryonic suppression. In Sec.~\ref{sec:results} we present the main results of the paper, i.e., the effects of baryons on the weak-lensing peak statistic. Finally, we conclude our work in Sec.~\ref{sec:conclusion}. Further details about the box replication method and resolution effects are found in the appendices \ref{sec:replication_and_randomization} and \ref{sec:mass_resolution}.

\section{Numerical Methods}
\label{sec:numerical_methods}
In this section we describe the process by which we generate weak-lensing convergence maps using $N$-body simulations and we discuss the angular power spectrum and the peak count statistic for the gravity-only case without baryonic effects. The convergence maps are based on particle light-cones which we construct by replicating standard $N$-body simulation outputs at many redshifts. In total, we generate fifty full-sky convergence maps for each survey configuration based on a set of ten statistically independent $N$-body simulations. This is possible because we randomise the particle positions in the replication process, generating five different convergence maps for every $N$-body simulation. In order to emulate different types of surveys with various numer densities, we furthermore post-process the generated convergence maps using different noise levels and smoothing scales.

\subsection{\textit{N}-body Simulations}
\label{sec:n_body_simulations}

The $N$-body simulations are generated using the publicly available $N$-body code {\tt PKDGRAV3} \citep{pkdgrav,pkdgrav3}. We assume a flat $\Lambda$CDM cosmology with best-fitting parameters from the Planck 2015 \citep{planck_collaboration_2015} release (see Table~\ref{tab:cosmology}). The initial conditions are generated with the {\tt MUSIC} code \citep{music} assuming a transfer function based on the fitting function from \citet{eisenstein_1998}.
\begin{table}[h]
\centering
\begin{tabular}{c | c | c | c | c | c}
$h$ & $n_s$ & $\Omega_\Lambda$ & $\Omega_M$ & $\Omega_b$ & $\sigma_8$\\\hline
0.67 & 0.96 & 0.68 & 0.32 & 0.045 & 0.83 \\
\end{tabular}
\caption{Set of cosmological parameters used in all our $N$-body simulations.}
\label{tab:cosmology}
\end{table}
We simulate sub-volumes of box-length $L=512\,\, \text{Mpc}/h$ each with $N=512^3$ particles which are then replicated to fill the full survey volume. The particle positions within a simulation volume are offset by a random vector and the volumes themselves rotated during the replication process in an attempt to reduce any repetitive patterns that may appear due to the large number of replications. The justification behind this replication and randomization process and its effects are discussed in more detail in Appendix \ref{sec:replication_and_randomization}.

Given this configuration and set of cosmological parameters, our simulations have a particle mass resolution of $M_\text{particle} = 9 \times 10^{10}\, M_\odot$/h. For the purpose of detecting the dark matter haloes in our \textit{N}-body simulations we use AMIGA Halo finder \citep{ahf}. We only select haloes with over 100 particles which corresponds to a halo mass resolution of $M_\text{halo} = 9 \times 10^{12}\, M_\odot$/h. In Appendix \ref{sec:mass_resolution} we show that our resolution is sufficient to obtain converged results for the weak-lensing peak count.

\subsection{Convergence Maps}\label{sec:convergence_maps}
We use a similar procedure as in Refs. \cite{teyssier_2009,sgier_2018} to generate weak-lensing convergence maps for a single non-tomographic redshift bin of the size $z=0.1-1.5$. The lightcone is constructed using 78 concentric spherical shells, with each shell being filled by the particles of the replicated simulation volume using the respective snapshot of the \textit{N}-body simulation. The simulated sub-volumes are replicated to fill the full survey volume of $\sim6\,\, \text{Gpc}/h$ using up to 1728 replications. The particles within each shell are projected onto a HEALPix map (of resolution $N_{\rm side}=4096$), weighted according to a realistic galaxy source distribution and summed up along the line-of-sight assuming the Born approximation.  The equation for the convergence of a single pixel on the HEALPix map can be formulated as:
\begin{equation}
\label{eq:convergence_map_kappa_pixelized}
\kappa(\hat{\theta}) = \frac{3}{2} \frac{H_0^2\Omega_m }{c^2}\sum_b W_b \left( \frac{12\cdot \text{NSIDE}^2}{4\pi}\frac{L^3}{N}\frac{n_b(\hat{\theta})}{\chi(z_b)^2} - \left[\chi(z_b + \Delta z_b) - \chi(z_b - \Delta z_b)\right] \right),
\end{equation}
where $n_b(\hat{\theta})$ is the particle count within the $b$-th shell at position $\hat{\theta}$ and $\Delta z_b$ is the shell width at redshift $z_b$. The shell weights $W_b$ are defined as
\begin{equation}
\label{eq:convergence_map_kappa_weights}
W_b = \left(\int_{\Delta z_b}\frac{\text{d}z}{E(z)}\frac{\chi(z)}{a(z)}g(z)\right) \Big/ \left(\int_{\Delta z_b}\frac{\text{d}z}{E(z)}\int_z^{z_H} \text{d}z' n_s(z')\right),
\end{equation}
where $\chi$ is the comoving distance. The lens efficiency $g(z)$ and the source distribution $n_s(z)$ are given by
\begin{equation}
g(z)=\frac{3}{2}\Omega_m\left(\frac{H_0}{c}\right)^2(1+z)\int_{z}^{z_\text{max}}n_{\rm s}(z)\frac{\chi(z')-\chi(z)}{\chi'(z)}dz'
\end{equation}
\begin{equation}
\label{eq:convergence_map_sources}
n_s(z)  \propto  z^2  \text{exp}\left(-\frac{z}{0.24}\right) 
\end{equation}
assuming a maximum redshift of $z_\text{max}=1.5$. The galaxy source distribution in Eq. \eqref{eq:convergence_map_sources} is a reasonable approximation for modern galaxy surveys.

After generating the raw convergence maps, we have to further account for the noise that is introduced in the observation process. For a detailed discussion of the noise and smoothing in convergence maps we refer to Ref.~\citep{vanwaerbeke_2000}. The pixelization of the convergence field can be approximated as smoothing using a top-hat filter. The Gaussian noise distribution is then described by
\begin{equation}
\label{eq:noise_tophat}
\langle\sigma_{\kappa,\text{pix}}^2\rangle = \frac{\sigma_e^2}{A_\text{pix}\,n_\text{gal}},
\end{equation}
where $A_\text{pix}$ is the area of a single pixel, $\sigma_e$ is the root mean square of the intrinsic ellipticity dispersion of galaxies  and $n_\text{gal}$ the galaxy number density. We add Gaussian smoothing to reduce the noise and to better mimic realistic convergence maps generated from galaxy surveys. The noise after Gaussian smoothing is given by
\begin{equation}
\label{eq:noise_gaussian}
\langle\sigma_\kappa^2\rangle = \frac{\sigma_e^2}{4\pi\theta_\kappa^2\,n_\text{gal}},
\end{equation}
where $\theta_k$ represents the smoothing scale. The choice of the smoothing scale determines the level of noise, with a trade-off that larger smoothing scales lead to a loss of small-scale information.

In order to quantify the expected difference beteween current and future galaxy surveys, we assume two separate survey configurations for all our convergence maps. We generate a measurement by first adding noise corresponding a given galaxy number density, before smoothing with an arbitrary smoothing scale. From each full-sky map, we then take multiple measurements by cutting out multiple (non-overlapping) circular patches, where the size of each patch corresponds to the survey area. Any additional systematic effects due to the geometry of the survey footprint are ignored. Since the smoothing and noise are added before any patches are created, this should not affect the measurements of peaks.

The first survey configuration, which we will refer to as \emph{DES-like}, assumes an area of $5000\  \textrm{deg}^2$ and a galaxy number density of $n_\text{gal} = 6\ \text{arcmin}^{-2}$, allowing us to cut six independent patches from each full-sky map. The second survey configuration, which we will refer to as \emph{Euclid-like}, has an area of $20000\ \textrm{deg}^2$ and a galaxy number density of $n_\text{gal} = 30\ \text{arcmin}^{-2}$. This setup only allows the use of 2 independent patches for each full-sky map.

For both configurations we use an ellipticity dispersion of $\sigma_e = 0.3$. We then smooth the resulting noisy maps assuming four different smoothing scales
\begin{equation}
\label{eq:map_smoothing}
\theta_\kappa \in \{2,\, 4,\, 8,\, 16\}\ \text{arcmin}.
\end{equation}

\subsection{Power Spectrum}
\label{sec:power_spectrum}
Two-point statistics correspond to the standard way of measuring the cosmological clustering process. Regarding the particle outputs from $N$-body simulations, we rely on the matter power spectrum
\begin{equation}
P(k, z) \propto \langle \tilde{\delta}(k, z) \tilde{\delta}^*(k, z)\rangle,
\end{equation} 
where $\tilde{\delta}(k, z)$ are the matter perturbations in Fourier space. For the case of two-dimensional convergence maps, we use the angular power spectrum
\begin{equation}
C(l) = \frac{1}{2l + 1}\sum_{m=-l}^l q_{lm}q_{lm}^*,
\end{equation}
with the coefficients $q_{lm}$ representing the spherical harmonic basis functions of the decomposed convergence map. These two measures are not independent and can be directly related with each other using the Limber approximation \citep{kilbinger_2015}: 
\begin{equation}
\label{eq:limber}
C(l) = \int_{0}^{\chi(z_\text{max})} g^2(\chi(z))P\left(\frac{l}{\chi}, z(\chi)\right)d\chi.
\end{equation}
The Limber approximation allows us to evaluate the accuracy of our generated maps by comparing the angular power spectrum directly measured on the convergence maps with the one obtained via Eq.~(\ref{eq:limber}) using the matter power spectrum from the \textit{N}-body simulations.

\begin{figure}[tbp]
  	\centering
	\includegraphics[width=.98\textwidth,trim=1.6cm 0.4cm 0.8cm 1.8cm,clip]{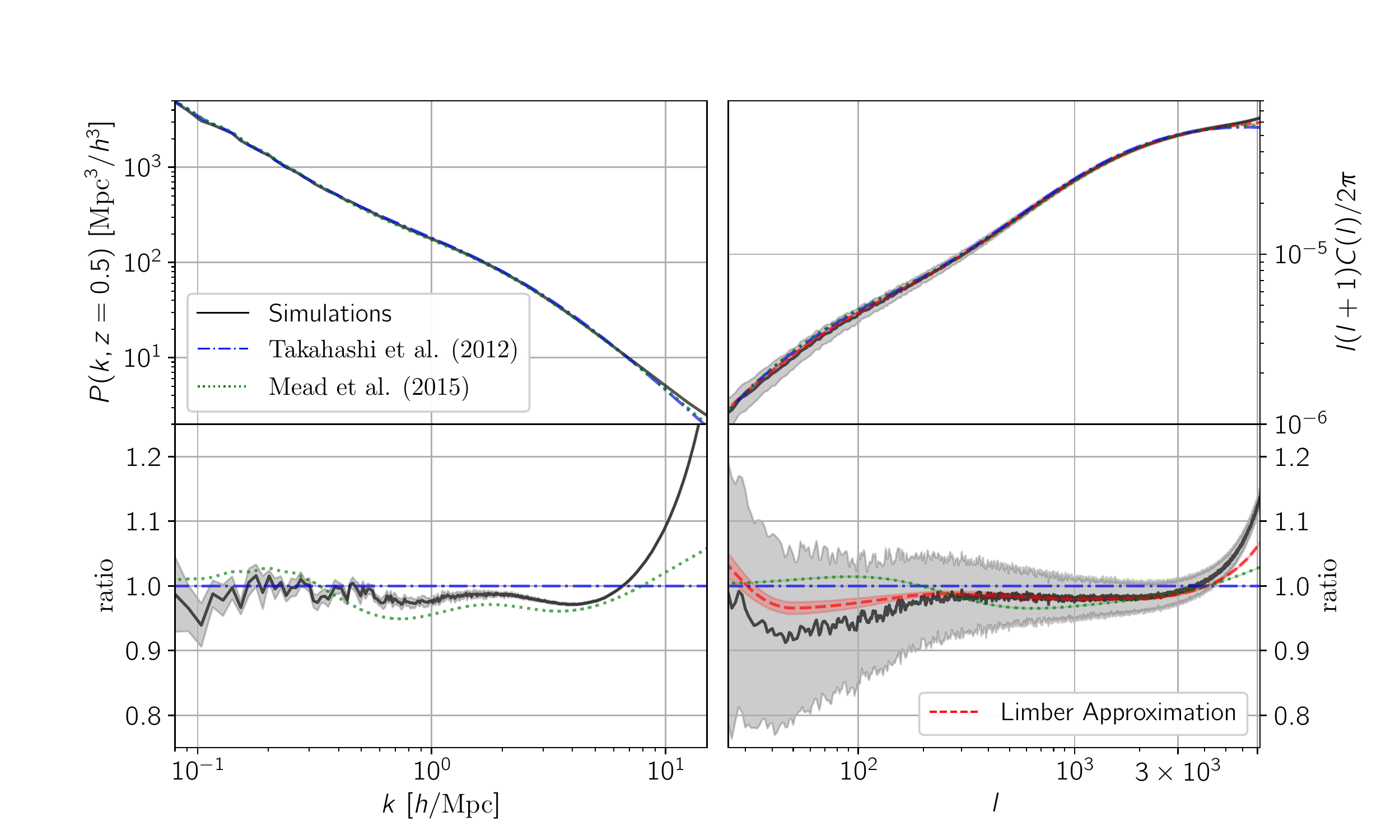}
	\caption{\emph{Left:} Average matter power spectrum from our dark-matter-only $N$-body simulations (black solid line) compared to the predictions from Takahashi et al. \citep{takahashi_2012} and Mead et al. \citep{mead_2015} (blue and green lines). The grey shaded band shows the spread from the ten different $N$-body runs. \emph{Right:} Average angular power spectrum directly measured from the convergence maps (black solid line) and calculated from the matter power spectrum via the Limber approximation (red dashed line), both with a surrounding band showing the spread of the different simulations. The angular power spectra from Takahashi et al. and Mead et al. obtained via the Limber approximation are added as blue and green lines.}
	\label{fig:pkcl_dmo}
\end{figure}

In the left panel of Fig.~\ref{fig:pkcl_dmo} we show the average power spectrum measured from all our simulations at $z=0.5$ and compare it to the \citet{takahashi_2012} and \citet{mead_2015} semi-analytical results. The agreement is within a few percent except at very large scales (small $k$-modes), where our simulations are missing power due to the limited box-size effect (see Appendix \ref{sec:replication_and_randomization} for more information).

The panel at right-hand-side of Fig.~\ref{fig:pkcl_dmo} illustrates the angular power spectrum directly measured at the map level and compares it to the results from the Limber approximation based on the same underlying simulations. The two approaches show a very good agreement, validating our pipeline for generating weak-lensing convergence maps. The differences at lower multipoles can be attributed to the replication and randomization procedure which suppresses modes at scales comparable to the simulation box-length and above. We again refer to the Appendix for more details about resolution and replication effects, including a short discussion on why the systematical effects visible in Fig.~\ref{fig:pkcl_dmo} do not affect the main results of the paper.

\subsection{Peaks}
\label{sec:peaks}
In the previous section, we used the angular power spectrum to validate our pipeline for generating weak-lensing maps from $N$-body simulations. As a next step we now turn our focus towards investigating peak statistics of the convergence field.  

We use the definition of a peak as a single pixel on our convergence map that has a higher convergence ($\kappa$) value than any of its eight neighbouring pixels. It is convenient to quantify the peak height with respect to the signal-to-noise ratio
\begin{equation}
\mathcal{S}/\mathcal{N}=\frac{\kappa}{\sqrt{\langle\sigma^2\rangle}}\,,
\end{equation}
where we assume a uniform Gaussian noise ($\sigma$) across the entire convergence map. The peaks are measured in a combination of linear and logarithmic bins. For the DES-like survey we use 12 linear (up to $\mathcal{S}/\mathcal{N}\approx 3.5$) and 6 logarithmic bins. The Euclid-like survey uses 20 linear (up to $\mathcal{S}/\mathcal{N}\approx 5.2$) and 8 logarithmic bins. The logarithmic binning at large $\mathcal{S}/\mathcal{N}$ values helps to reduce the statistical error at higher peaks, where the number of peaks drops off significantly. The outermost bin with the largest value of $\mathcal{S}/\mathcal{N}$ contains approximately 10 -- 20 peaks for both the DES-like and Euclid-like setup.

\begin{figure}[tbp]
	\centering
	\includegraphics[width=\textwidth,trim=1.2cm 0.1cm 0.0cm 1.2cm,clip]{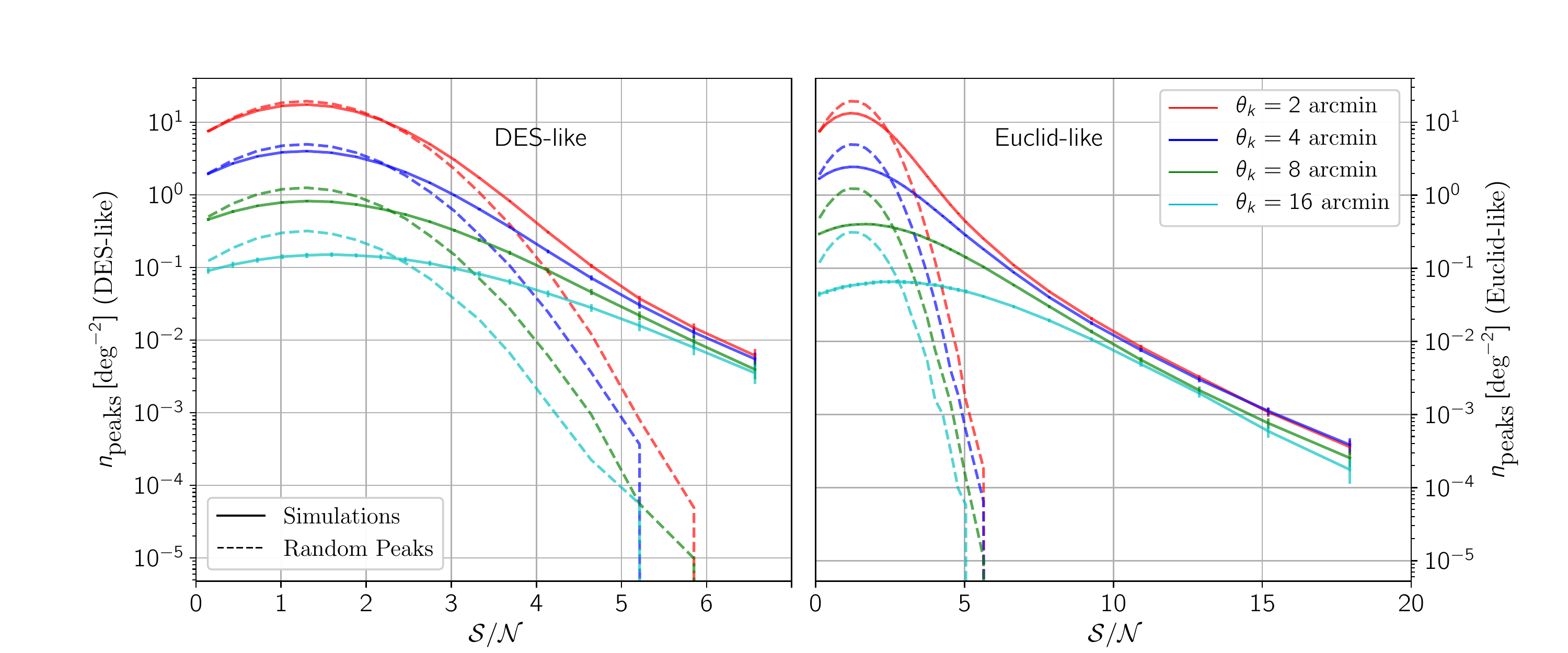}
	\caption{Peak densities of the generated convergence maps as a function of $\mathcal{S}/\mathcal{N}$ (solid lines). The peak densities from noise-only maps are shown for comparison (dashed lines). \emph{Left:} DES-like survey configuration using an area of 5000 $\textrm{deg}^2$ with a galaxy number density of $n_\textrm{gal} = 6\, \textrm{arcmin}^{-2}$. \emph{Right:} Euclid-like survey configuration using an area of 20000 $\textrm{deg}^2$ with a galaxy number density of $n_\textrm{gal} = 30\, \textrm{arcmin}^{-2}$.}
	\label{fig:peaks_dmo}
\end{figure}

Fig.~\ref{fig:peaks_dmo} illustrates the peak density as a function of $\mathcal{S}/\mathcal{N}$ for both DES-like and Euclid-like survey configurations and all four smoothing scales. Also shown are the peak densities from pure noise maps. The latter dominates the lower peaks.

Instead of counting the total number of peaks, it can also be convenient to define the noise-subtracted peak density given by
\begin{equation}
\label{eq:peaks_diff}
\Delta n_\textrm{peaks} = n_\textrm{peaks} - n_\textrm{peaks,\,noise},
\end{equation}
where $n_{\rm peaks, noise}$ is the peak-count from the noise-only map. This modified statistics allows us to immediately see where the peak counts are dominated by noise and how the cosmological signal compares to the size of the statistical error. 

\begin{figure}[tbp]
	\centering
	\includegraphics[width=\textwidth,trim=1.2cm 0.1cm 0.0cm 1.2cm,clip]{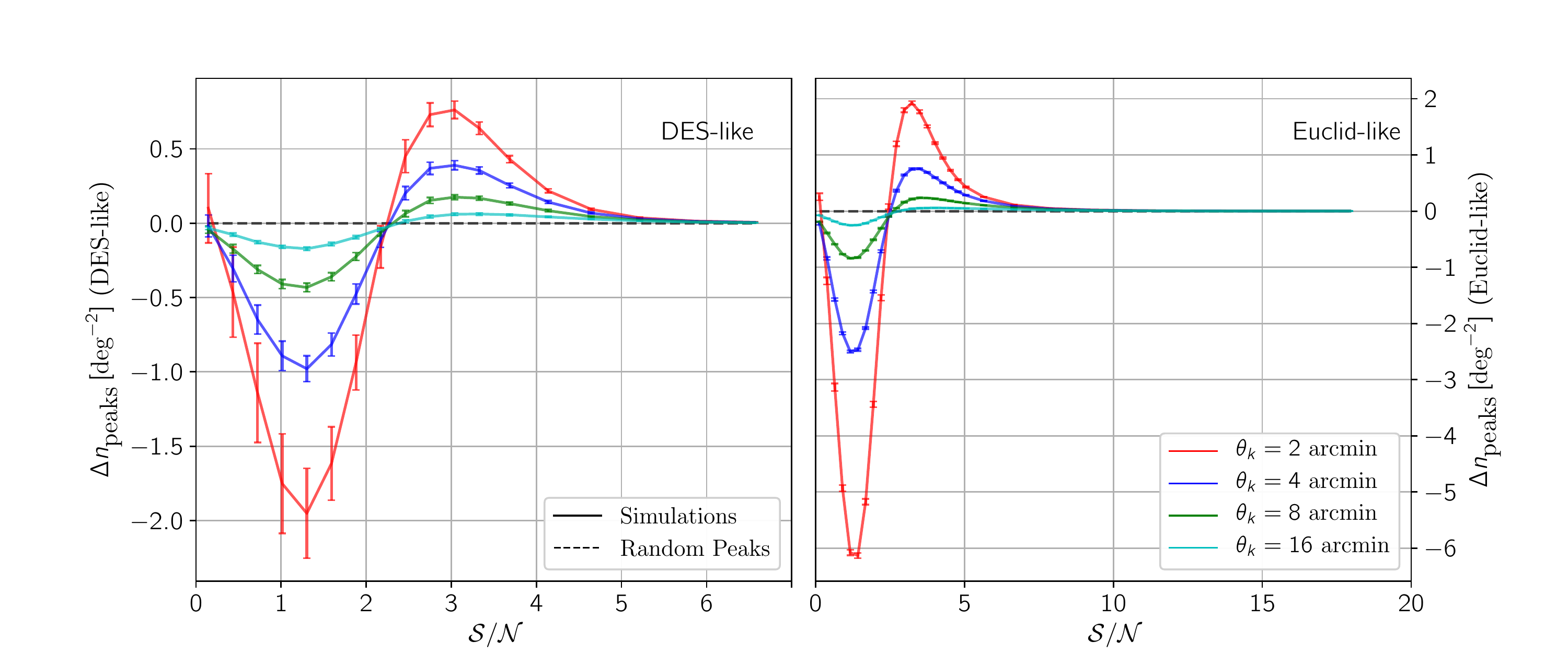}
	\caption{Same setup as in Fig.~\ref{fig:peaks_dmo}, except that noise peak density is subtracted from the simulated peak density (see Eq. \ref{eq:peaks_diff}). The vertical error bars show the statistical error for each bin, normalized with respect to the bin width, that is estimated from all measured patches (2 or 6 patches from 50 full-sky maps).}
	\label{fig:peaks_dmo_diff}
\end{figure}

In Fig.~\ref{fig:peaks_dmo_diff} we show the noise subtracted peak density function for various smoothing scales and both survey configurations. The results based on DES-like survey are in good qualitative agreement with previous work \citep[see Refs.][]{kacprzak_2016, fluri_2018}. A comparison between the two survey configurations highlights the expected improvement in accuracy that can be expected from future surveys. The four-times larger survey area and the five-times better galaxy number density of the Euclid-like configuration compared to the DES-like setup results in a threefold increase of the signal in the Gaussian noise subtracted peaks. This means that future weak-lensing surveys will be capable of detecting small deviations of the peak distribution and therefore require a careful treatment of all systematics. This is especially true regarding the effects of baryons as we will see in the following sections.

\section{Baryonic Correction Model}\label{sec:baryonic_correction_model}
The baryonic correction (BC) model consists of a method to modify outputs of $N$-body simulations in order to mimic the effects of baryons on the cosmological density field. Here we provide a summary of the method and the parametrisation. More details can be found in ST15 \citep{schneider_2015} and S19 \citep{schneider_2018}.

The BC model is based on an algorithm to radially displace particles around halo centres in $N$-body simulations. The goal is to slightly perturb individual halo profiles in order to account for baryonic effects without losing the information regarding the full cosmological density field. The displacement of particles relies on a parametrisation of the halo profile which includes both dark matter and baryonic components, and which is motivated by observations. The halo profile is given by
\begin{equation}\label{rhotot}
\rho_{\rm halo}(r)=\rho_{\rm gas}(r) + \rho_{\rm cga}(r) + \rho_{\rm clm}(r),
\end{equation}
where the different terms refer to the gas (gas), the central galaxy (cga), and the collisionless matter (clm) components. The latter consists of both dark matter (dm) and the satellite galaxies (sga) of the halo. In the following, we provide a short summary of the parametrisation:

\begin{itemize}
\item The gas profile ($\rho_{\rm gas}$) is described by a power law with inner core and outer truncation, motivated by stacked X-ray observations ~\citep[see for example Refs.][]{Eckert:2012aaa,Eckert:2015rlr}. It is given by
\begin{equation}\label{rhogas}
\rho_{\rm gas}(r)\propto\left[1+\frac{r}{r_{\rm co}}\right]^{-\beta} \left[1+\left(\frac{r}{r_{\rm ej}}\right)^2\right]^{(7-\beta)/2},
\end{equation}
where the core and ejection radii are fixed to $r_{\rm co}=0.1\times r_{200}$ and $r_{\rm ej}=4\times r_{200}$, for simplicity. We refer to S19 for a more general parametrisation and a discussion about the effects of varying $r_{\rm co}$ and $r_{\rm ej}$ on the cosmological density field. The slope of the gas profile is given by the equation
\begin{equation}\label{beta}
\beta=3-\left(\frac{M_{\rm c}}{M_{\rm 200}}\right)^{\mu},
\end{equation}
imposing a decreasing value of $\beta$ towards smaller halo masses, again motivated by X-ray observations \citep[][]{Eckert:2015rlr}. The free model parameters $M_c$ and $\mu$ control the pivot scale and the slope of this mass dependence.

\item The central galactic profile ($\rho_{\rm cga}$) is described in the baryonic correction model as a simple power law that is exponentially truncated beyond the half-light radius. Note, however, that the stellar profile only affects small scales and has therefore little importance for the weak-lensing signal.

\item The collisionless matter profile ($\rho_{\rm clm}$) is given by a truncated NFW profile which is assumed to undergo adiabatic relaxation. The latter leads to a steepening of the profile in the centre and a flattening in the outskirts induced by the stellar and the gas components.
\end{itemize}
Note that the BC model only slightly perturbs the cosmological density field without destroying its complex structure, which includes voids, sheets, filaments, and haloes of triaxial shape. However, it does induce a flattening of profiles, a change in virial mass of haloes, and a slight displacement of halo positions in agreement with expectations from baryonic physics. The fact that the BC model only slightly perturbs the nonlinear matter distribution makes it much more realistic than more analytical approaches such as the halo model, where basic assumptions of randomly distributed halo positions or perfectly symmetric halo profiles have to be made.

The version of the BC model sketched out above has two free model parameters ($M_c$, $\mu$) regulating the amount of gas inside and outside of haloes. We now follow S19 and use observed gas fractions from X-ray data to constrain these parameters.

An important problem regarding gas fractions from X-ray observations is that they are derived assuming hydrostatic equilibrium of the gas. We account for this potential systematic by including a hydrostatic mass bias
\begin{equation}\label{hsbias}
1-b_{\rm hse}\equiv \frac{M_{\rm 500,hse}}{M_{500}},
\end{equation}
where $M_{500}$ is the actual halo mass at $r_{500}$, while $M_{\rm 500,hse}$ is the mass obtained from X-ray observations under the assumption of hydrostatic equilibrium. Results from hydrodynamical simulations suggest the hydrostatic bias ($b_{\rm hse}$) to be between 0 and 40 percent. Based on this, we define three \emph{benchmark models} that correspond to an optimistic, best-guess, and pessimistic scenario with $1-b_{\rm hse}=1.0$, $0.83$, $0.71$, respectively. 

\begin{figure}[tbp]
\center{
\includegraphics[height=.36\textwidth,trim=-0.5cm 0.1cm 1.2cm 0.8cm,clip]{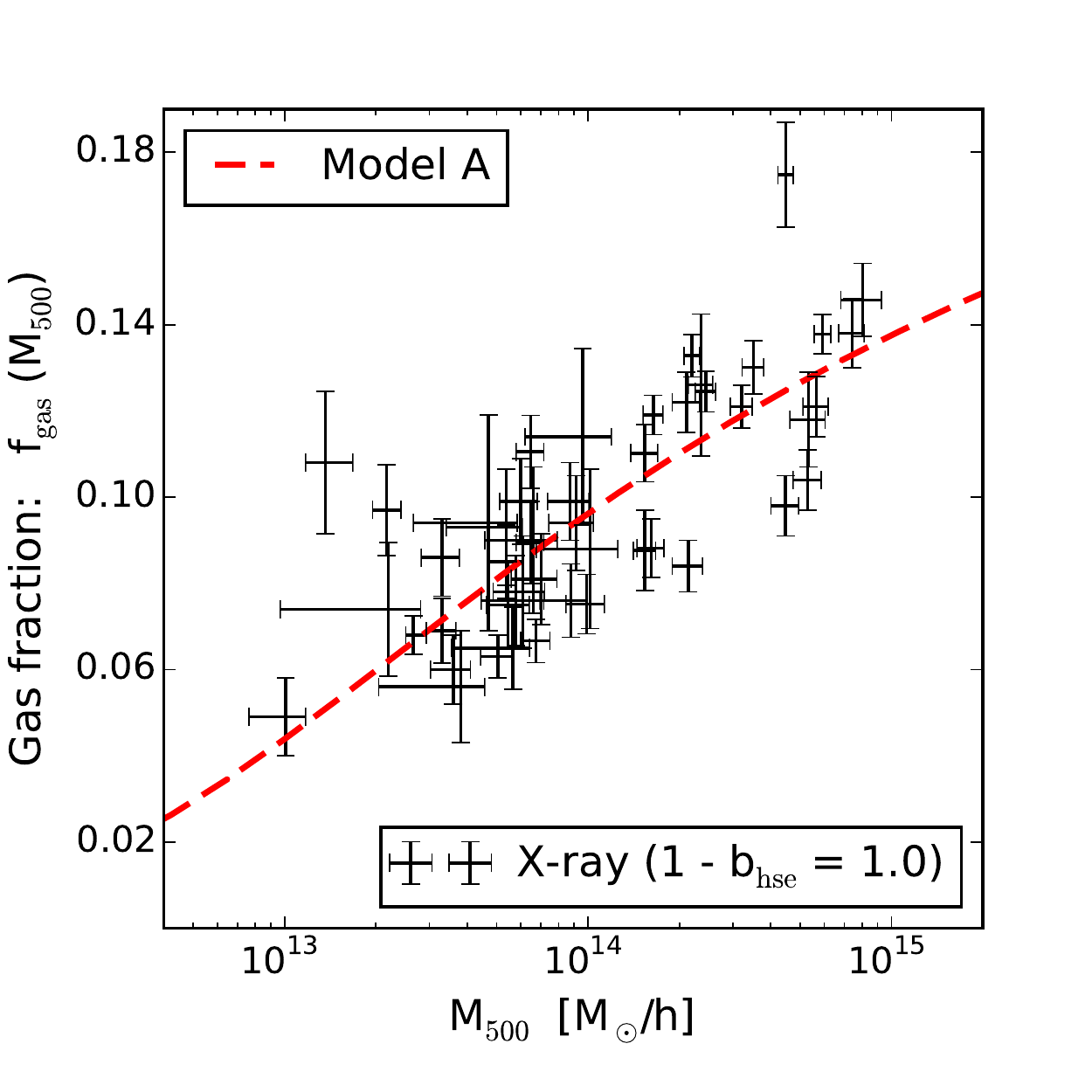}
\includegraphics[height=.36\textwidth,trim=1.85cm 0.1cm 1.2cm 0.8cm,clip]{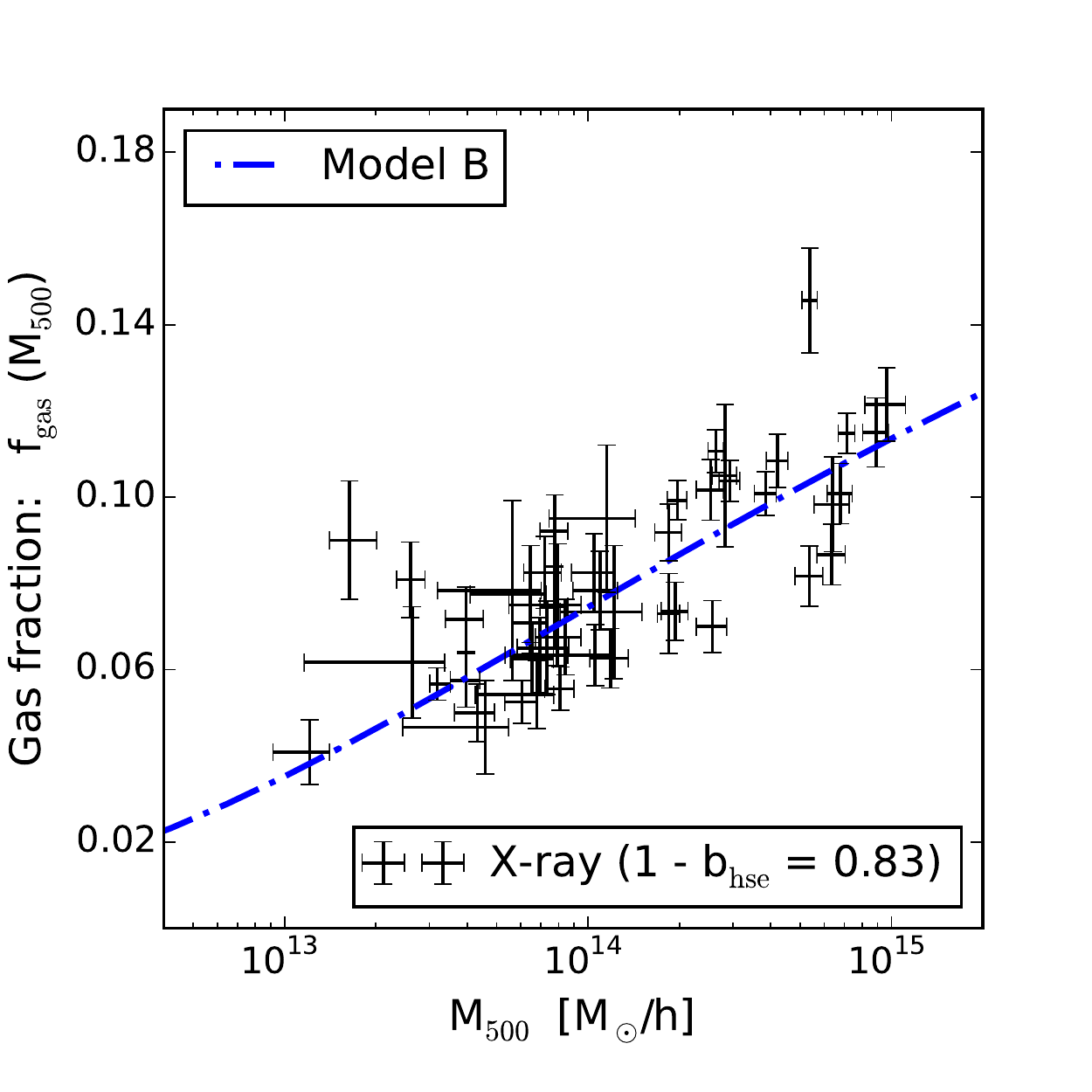}
\includegraphics[height=.36\textwidth,trim=1.85cm 0.1cm 1.2cm 0.8cm,clip]{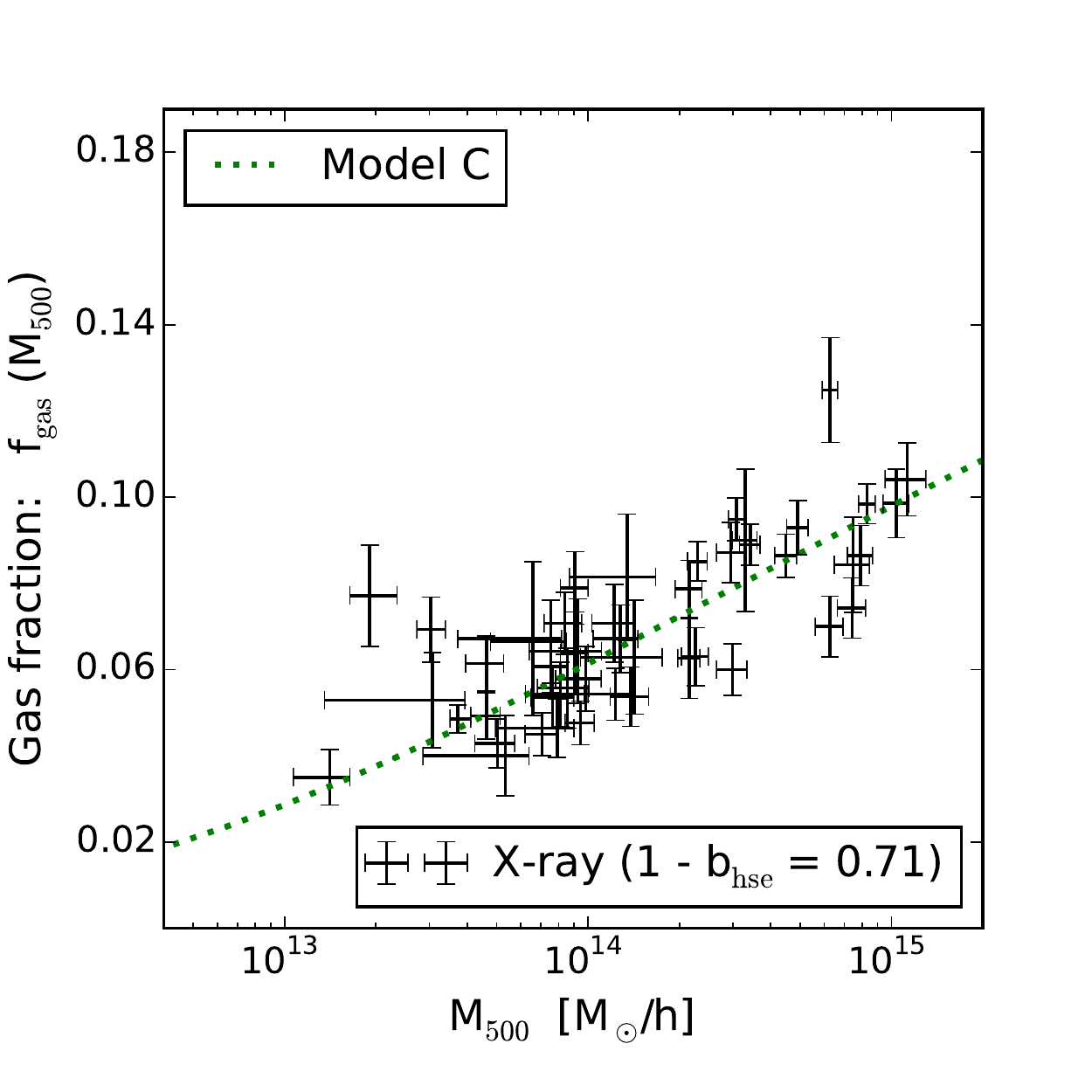}\\
\includegraphics[width=.99\textwidth,trim=1.58cm 0.1cm 2.78cm 1.3cm,clip]{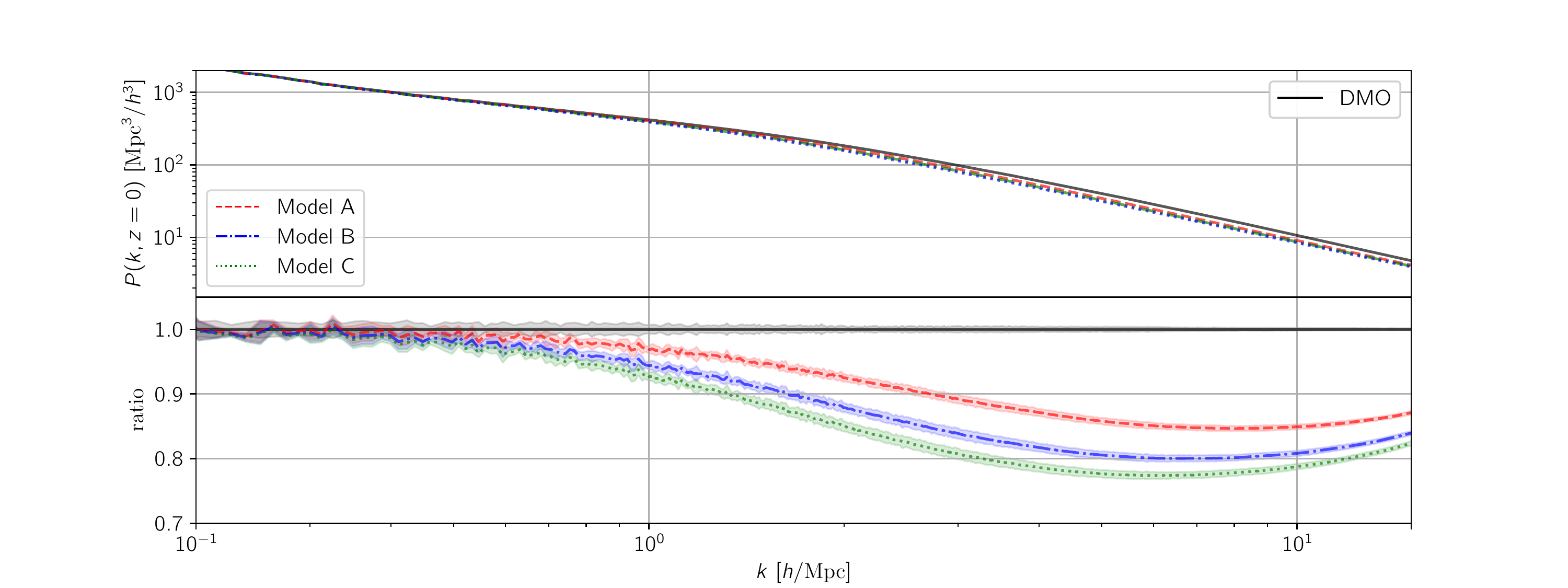}
\caption{\label{fig:PSvarparams}Calibration of the BC model with X-ray data and resulting matter power spectra. \emph{Top:} fraction of gas to total matter ($f_{\rm gas}$) at $r_{500}$ from a selection of X-ray observations~\citep{Sun:2008eh,Vikhlinin:2008cd,Gonzalez:2013awy} (black symbols) assuming a hydrostatic bias of $1-b_{\rm hse}=1.0$, $0.83$, $0.71$ (left to right). The coloured lines show the average gas fraction from the BC model, where the parameters ($M_c$, $\mu$) have been fitted to the different X-ray data sets. The three cases correspond to the benchmark models A, B, and C representing an optimistic, a best-guess, and a pessimistic scenario for the baryonic suppression effect. \emph{Bottom:} The resulting power spectra of the three benchmark models. The coloured band around the lines show the spread in the predictions from ten different $N$-body simulations.}}
\end{figure}

The top panels of Fig.~\ref{fig:PSvarparams} show the gas fractions from a collection of X-ray observations \citep{Sun:2008eh,Vikhlinin:2008cd,Gonzalez:2013awy} assuming the three different values for the hydrostatic mass bias mentioned above. This means that the first panel shows the uncorrected gas fractions from the literature, while the second and third panel is corrected to account for a hydrostatic mass bias of 20 and 40 percent, respectively. A large hydrostatic mass bias leads to an under-estimation of the total halo mass (see Eq.~\ref{hsbias}). Correcting for this causes the data points in the upper-central and upper-right panel to shift further downwards and to the right compared to the data points in the upper left panel. A more detailed explanation of the connection between X-ray gas fractions and the hydrostatic mass bias can be found in S19 \citep{schneider_2018}. Next to the X-ray data points, the top panels show the gas fractions obtained with the BC model where, the model parameters ($M_c$, $\mu$) are fitted to the data. This fitting procedure defines the BC models A, B, and C that we will used as benchmark scenarios throughout this paper. The characteristics of the benchmark models are summarised in Table \ref{tab:benchmark}. The values of the benchmark parameters are listed in Table~\ref{tab:benchmark}. 

The bottom panel of Fig.~\ref{fig:PSvarparams} shows the resulting power spectra of the three benchmark models. They feature a maximum suppression of about 15, 20, and 25 percent, repsectively. The curves agree with the power spectra of models A-avrg, B-avrg, and C-avrg from S19 to better than one percent for $k<5$ h/Mpc. The reason why they are not identical is the underlying $N$-body simulations which have a larger volume but lower resolution compared to the ones used in S19. We refer to Appendix \ref{sec:mass_resolution} for a detailed discussion about resolution effects.

Throughout this paper, we will use the benchmark models defined above to quantify the range of realistic baryonic correction effects. They stand for an optimistic (model A), a best-guess (model B), and a pessimistic (model C) scenario regarding the baryonic effects on the cosmological density field. This characterisation is reasonable because the hydrostatic mass bias has been identified in S19 to be the largest systematic uncertainty of the analysis pipeline.

\begin{table}
\caption{Baryonic benchmark models with parameters ($M_c$, $\mu$) obtained by fitting the baryon fraction of the BC model to X-ray observations. Models A, B, and C are based on different assumptions regarding the hydrostatic-mass bias ($b_{\rm hse}$). The gas ejection radius has been fixed to $\theta_{\rm ej}=4$ in agreement with observed gas profiles.}
\label{tab:benchmark}
\begin{center}
\small
 \setcellgapes{1pt}\makegapedcells
\begin{tabular}{p{2.0cm}p{1.5cm}p{0.5cm}p{2.0cm}p{0.6cm}p{6.0cm}}
\hline
 Name  & $1-b_{\rm hse}$ & $\theta_{\rm ej}$ & $M_{\rm c}$ [M$_{\odot}$/h] & $\mu$  & Baryonic scenario \\
 \hline
 \hline
 Model A & 1.000 & 4 &$2.3\times10^{13}$ & 0.31 & optimistic (weak baryonic effects)\\
 
 Model B & 0.833 & 4 & $6.6\times10^{13}$ & 0.21 & best-guess (medium baryonic effects)\\
 
 Model C & 0.714 & 4 &$1.9\times10^{14}$ & 0.17 & pessimistic (strong baryonic effects)\\
 \hline
\end{tabular}
\end{center}
\end{table}

Before moving on to the result section, it is important to emphasise that, although consisting of an approximative method, the BC model has been shown to be in good agreement with full hydrodynamical simulations. As a test, S19 calibrated the BC model parameters using the measured gas fractions from hydrodynamical simulations at redshift zero. They then predicted the matter power spectrum based on the BC model (using the fitted parameters) and compared it with the power spectrum of the same hydrodynamical simulation. At redshift zero the agreement between the BC model and the simulations was better than two percent for all modes below $k=5$ h/Mpc. At redshift one and two the agreement degraded slightly but was still better than 3 and 5 percent, respectively. The decrease in accuracy towards higher redshift is not surprising since the model parameters have been calibrated against gas fractions at redshift zero. The comparison of S19 included the hydrodynamical simulations OWLS \citep{schaye_2010,vandaalen_2011}, cosmo-OWLS with AGN8.0 and AGN8.5  \citep{lebrun_2014,mummery_2017}, Horizon-AGN \citep{dubois_2016,chisari_2018}, and Illustris-TNG \citep{springel_2018}.

\section{Results}
\label{sec:results}
With the BC model introduced in the previous section, it is straight-forward to include baryonic effects into the pipeline for weak lensing convergence maps. We simply apply the particle displacement algorithm to all $N$-body outputs at different redshifts, before building the particle light-cone and producing the angular maps. For each of the ten $N$-body runs, we perform the baryonic correction three times with model parameters corresponding to the three benchmark models summarised in Table~\ref{tab:benchmark}.

\begin{figure}[tbp]
  	\centering
	\includegraphics[width=\textwidth, trim=1.74cm 0.18cm 2.5cm 1.2cm,clip]{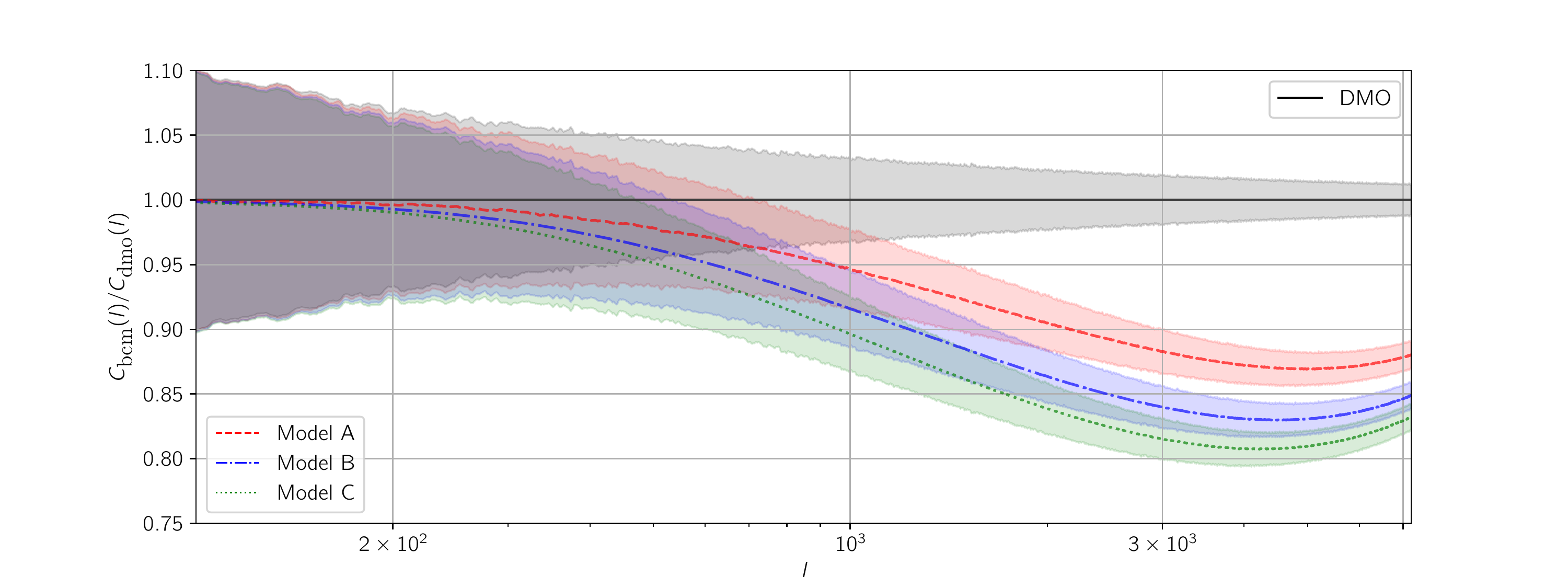}	
	\caption{Ratio between the angular power spectrum with and without baryonic effects from the simulated weak-lensing convergence maps. The three benchmark models shown as coloured lines correspond to an optimistic, best-guess, and pessimistic baryonic suppression scenario (see Sec.~\ref{sec:baryonic_correction_model}). The shaded bands around the lines represent the sample variance obtained from all 50 maps.}
	\label{fig:cl_baryons}
\end{figure}

As a first step we investigate the effects of baryons on the angular power spectrum measured from the convergence maps. Fig.~\ref{fig:cl_baryons} illustrates the ratio between the angular power spectrum with and without baryons ($C_{\rm bcm}$ and $C_{\rm dmo}$, respectively). Depending on the benchmark model, the baryonic effects lead to a suppression of up to 20 percent starting beyond $l\sim200-400$ and dominating the statistical error at  $l\sim500-800$. These scales are relevant for current weak-lensing surveys such as KiDS, DES, or HSC. Recent results from the last of these have shown measurements of the angular power spectrum up to $l\sim3000$ with error bars comparable to the amplitude of the baryon effects shown here \citep[see Ref.][]{hikage_2018}.

The results of Fig.~\ref{fig:cl_baryons} are generally consistent with the findings of S19 \citep{schneider_2018}. Note however, that compared to S19, the amplitude of the baryon suppression is slightly smaller by about 1-2 percent below $l\sim3000$ and $\sim5$ percent below $l\sim6000$. These differences can be attributed to the lower mass resolution of the simulations used here. See Appendix \ref{sec:mass_resolution} for a more detailed analysis of resolution effects.

We now turn our attention towards the main goal of the paper which is the measurement of baryonic effects on the weak-lensing peaks. A convenient statistical measure to determine the significance of the baryon effects ($s_{\rm bcm}$) is the relative difference of the peak distributions divided by the noise, i.e.,
\begin{equation}
s_{\rm bcm}\equiv\left(n_{\rm bcm}-n_{\rm dmo}\right)/\sigma_{\rm dmo}.
\end{equation}
A significance of $|s_{\rm bcm}|\ll1$ means that the baryonic effects are subdominant compared to the statistical error and can be ignored. On the other hand, for the case of $|s_{\rm bcm}|>1$, the baryonic effects are expected to dominate the error budget and have therefore to be properly included in the analysis pipeline.

\begin{figure}[tbp]
  	\centering	
	\includegraphics[width=\textwidth,trim=0.25cm 0.5cm 0.5cm 1.8cm,clip]{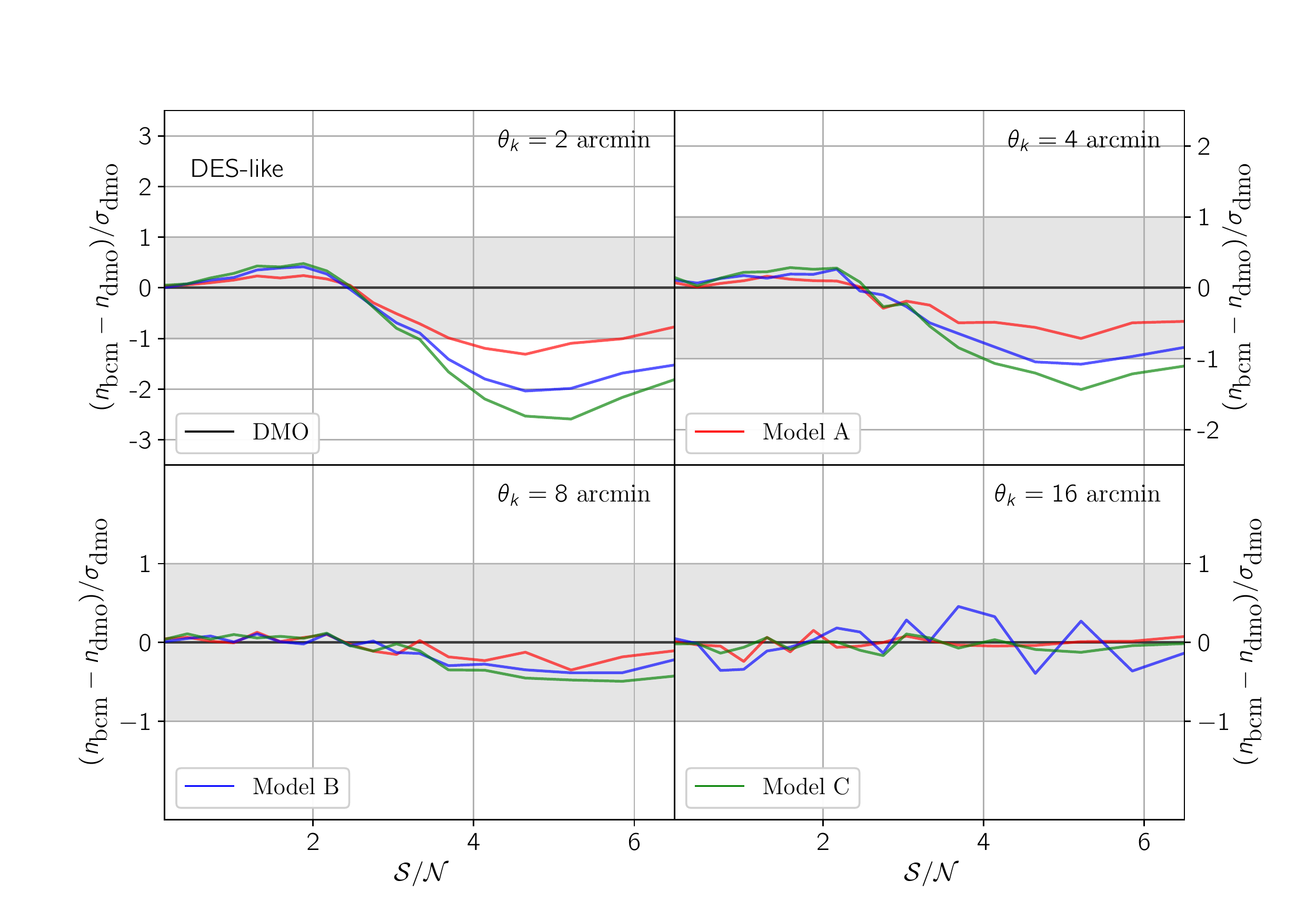}
	\caption{Significance of baryonic effects on the peak numbers of weak-lensing convergence maps assuming a DES-like survey configuration and four different smoothing scales ($\theta_k$). The grey band shows the expected level of noise ($\sigma_{\rm dmo}$). The three benchmark models (summarised in table \ref{tab:benchmark}) correspond to an optimistic, best-guess, and pessimistic baryonic suppression scenario (coloured lines).}
	\label{fig:peaks_baryons_des}
\end{figure}

Fig.~\ref{fig:peaks_baryons_des} illustrates the significance of the baryonic effects for the peak distributions of a DES-like survey configuration, where each panel refers to a different smoothing scale. Unsurprisingly, the smaller the smoothing scale the larger the significance of the baryonic feedback effects. A smoothing of $\theta_k=8$ arcmin or higher is sufficient to guarantee that baryonic effects stay subdominant over all scales of the DES configuration. This is not the case for smaller values of $\theta_k$, where the high peaks become significantly affected by baryons. For our most extreme case of $\theta_k=2$ arcmin, the baryons lead to a 1-2 $\sigma$ bias for scales above $\mathcal{S}/\mathcal{N}=3$. Smaller peaks below this threshold remain less affected by baryonic effects.

The results shown in Fig.~\ref{fig:peaks_baryons_des} validate the approach of \citet{kacprzak_2016}, a cosmological parameter study based on peak counts of the DES science verification data. While this paper did not include baryonic effects, it used a sufficiently large smoothing of the map to fully mitigate all potential systematics from baryonic feedback. The situation could be different regarding the more recent peak count analysis from KiDS \citep{shan_2018,martinet_2018} which applied a smoothing scale of  $\sim2$ arcmin and used both low and high signal-to-noise peaks. Our findings suggest that these results could indeed be affected by baryons at a significant level. Note, however, that more quantitative conclusions can only be obtained with a dedicated study that assumes a KiDS-like survey configuration in terms of mean galaxy number density, redshift distribution, and survey area.

\begin{figure}[tbp]
  	\centering
	\includegraphics[width=\textwidth,trim=0.25cm 0.5cm 0.5cm 1.8cm,clip]{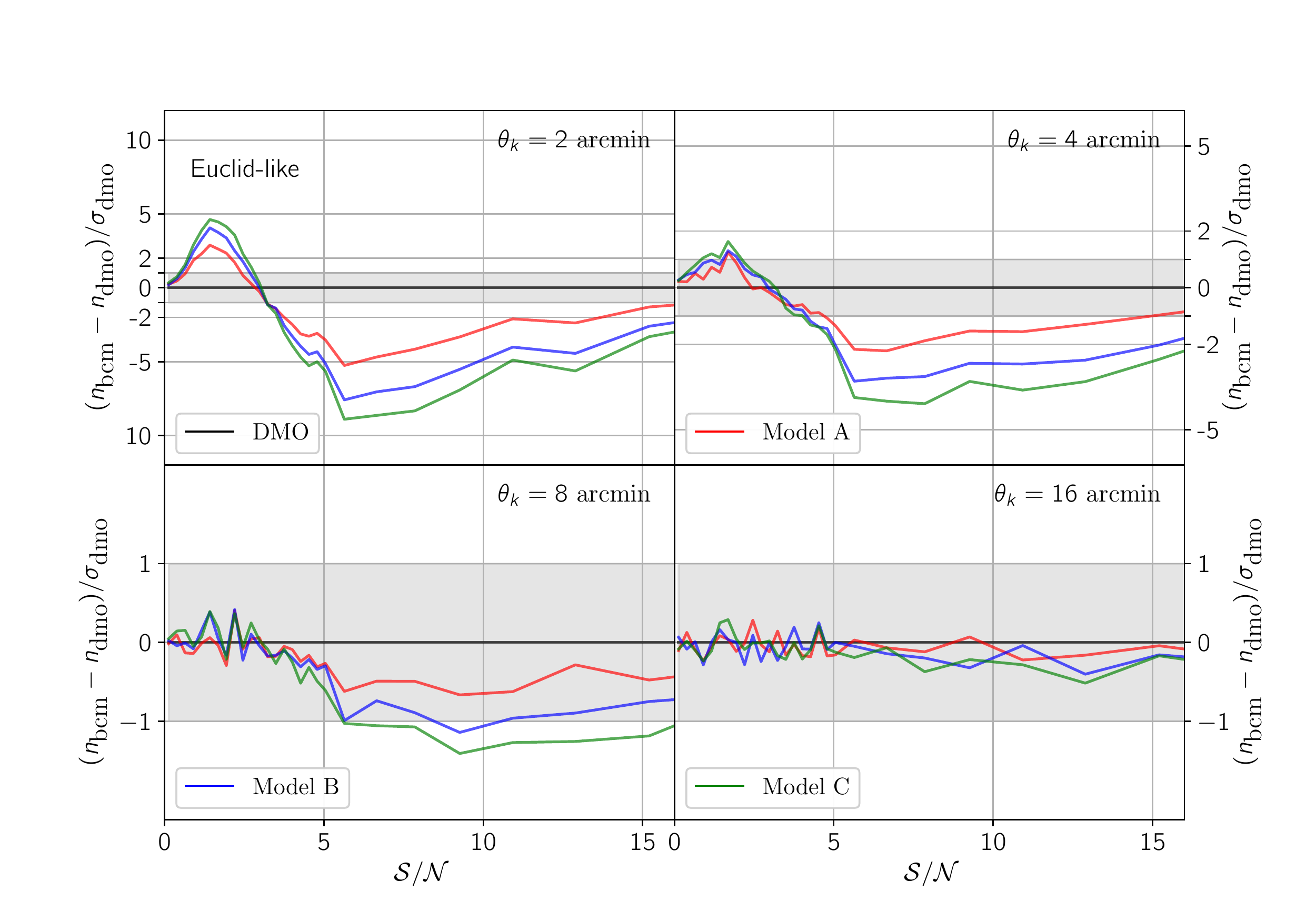}
	\caption{Significance of baryonic effects on the peak numbers of weak-lensing convergence maps assuming a Euclid-like survey configuration and four different smoothing scales ($\theta_k$). The grey band shows the expected level of noise ($\sigma_{\rm dmo}$). The three benchmark models (summarised in table \ref{tab:benchmark}) correspond to an optimistic, best-guess, and pessimistic baryonic suppression scenario (coloured lines). Note the significantly enhanced baryonic effects and the reduced statistical errors compared to Fig.~\ref{fig:peaks_baryons_des}.}
	\label{fig:peaks_baryons_euclid}
\end{figure}

In the future, baryonic feedback will become an important systematical effect for weak lensing surveys. In Fig.~\ref{fig:peaks_baryons_euclid} we show the same information as in Fig.~\ref{fig:peaks_baryons_des} but this time for a Euclid-like survey configuration. As expected, the strongly increased signal-to-noise leads to larger systematics from baryonic feedback. A smoothing of $\theta_k=16$ arcmin or above is required if baryonic effects are ignored in the peak modelling. However, such a large smoothing suppresses cosmological information and is therefore no viable option. At smaller smoothing scales below $\theta_k=8$ arcmin, baryonic feedback is significantly altering the signal. Regarding high peaks with $\mathcal{S}/\mathcal{N}>2$, we expect a bias of the order 2-4 $\sigma$ for $\theta_k=4$ arcmin and 5-8 $\sigma$ for $\theta_k=2$ arcmin. Furthermore, note that at these small smoothing scales the low peaks with $\mathcal{S}/\mathcal{N}<2$ also become affected at the 1-$\sigma$ level for $\theta_k=4$ arcmin and the 4-$\sigma$ level for $\theta_k=2$ arcmin.

While low $\mathcal{S}/\mathcal{N}$ peaks will also be influenced by baryons for future surveys like Euclid, the effect on high peaks remains clearly stronger.  We speculate that the reason for this behaviour is that high peaks are caused by individual haloes whose profiles are directly affected by baryonic feedback. Low peaks, on the other hand, are caused by a combination of multiple line-of-sight haloes and shape noise \citep{yang_2011,liu_2016}, where only the haloes are modified by baryonic feedback while the noise term stays the same.

Towards small smoothing scales, the relative peak counts of both DES and Euclid type surveys show a characteristic wave shape (see top panels of Fig.~\ref{fig:peaks_baryons_des} and \ref{fig:peaks_baryons_euclid}) with $n_{\rm bcm}>n_{\rm dmo}$ at high and $n_{\rm bcm}<n_{\rm dmo}$ at low values of $\mathcal{S}/\mathcal{N}$. Since the total number of peaks changes by less than one percent between the baryonic corrected and the dark matter-only sample, we conclude that baryonic effects cause peaks to simply shift towards lower $\mathcal{S}/\mathcal{N}$ values without destroying them.

Based on the analysis shown above we conclude that for current, stage-III weak-lensing surveys, the baryonic suppression effects are at most comparable to the statistical error bars. We therefore expect that ignoring baryonic feedback does not strongly bias cosmological parameter estimates. For future, stage-IV weak-lensing surveys, on the other hand, it will be essential to properly include baryon effects in the prediction pipeline. Most notably, this is true for small smoothing scales below $\theta_k\sim8$ arcmin which contain important information about cosmology.

\section{Conclusion}\label{sec:conclusion}
In this paper we have investigated the effects of baryonic feedback on peak number statistics of the weak-lensing convergence field. The convergence maps were constructed via light-cones of $N$-body simulations which have been integrated along the line-of-sight using appropriate lensing kernels and simple noise configurations. The effects of baryons were implemented using the baryon correction model of S19 which can be directly applied to {\it N}-body simulations and has previously been shown to provide an accurate prescription of the baryon suppression effects at scales relevant for weak lensing \citep{schneider_2018}. Following S19 we defined three benchmark baryonic models motivated by X-ray observations of gas fractions which provide a realistic range of scales for the baryonic suppression effects.

In a first step we validated our pipeline by measuring the angular power spectrum using fifty full-sky convergence maps based on light-cones of replicated $N$-body outputs. A comparison to results from the literature based on the Limber approximation shows good agreement regarding both the baryon suppression effect and the absolute power spectrum of the dark-matter-only case.

As a further step we investigated the convergence peak count, assuming two distinct configurations for the survey footprint, the mean galaxy density, and the noise implementation. The first configuration is motivated by the characteristics of DES, while the second is based on the expected setup of Euclid. This allowed us to quantify the baryonic feedback effects of both current stage-III and future stage-IV weak-lensing experiments.

Not surprisingly, we found that the strength of the baryon suppression on the peak statistics strongly depends on the scale of the Gaussian smoothing applied to the convergences maps. For the DES-like configuration, a smoothing scale of $\theta_k\sim 8$ arcmin or larger is sufficient to wash-out the baryonic effects so that the suppression signal becomes smaller than the statistical error from limited number counts. For the Euclid-like configuration, on the other hand, a smoothing of at least $\theta_k\sim 16$ arcmin is required to reduce the baryon suppression signal enough so that it falls below the statistical error.

Smaller smoothing scales than the ones mentioned above require a proper modelling of baryon suppression effects for the weak-lensing convergence map. Assuming a DES-like survey configuration and a Gaussian smoothing of $\theta_k=2$ ($\theta_k=4$) arcmin, the baryon effects become larger than the statistical error for high peaks with $\mathcal{S}/\mathcal{N}>3$ ($\mathcal{S}/\mathcal{N}>4$). Low peaks are also modified by baryon feedback, but this effect stays much smaller than the sample variance error due to the limited survey footprint. 

The consequences of baryon feedback are more serious for the case of a Euclid-like survey setup. Assuming a smoothing scale of $\theta_k=4$ arcmin or below, the baryon effects exceed the expected statistical errors over the entire range of $\mathcal{S}/\mathcal{N}$. This means that for Euclid, not only high but also low peaks are significantly affected by baryons and a proper modelling of baryon feedback effects becomes indispensable.

\section*{Acknowledgments}
We acknowledge support from the Swiss National Science Foundation. This includes the grants PZ00P2\_161363 and 200021\_169130.


\bibliographystyle{unsrtnat}
\bibliography{peaks}

\appendix
\section{Replication and Randomization Effects}
\label{sec:replication_and_randomization}

\begin{figure}[tbp]
	\centering
	\includegraphics[width=\textwidth,trim=1.5cm 0.52cm 1.3cm 0.65cm,clip]{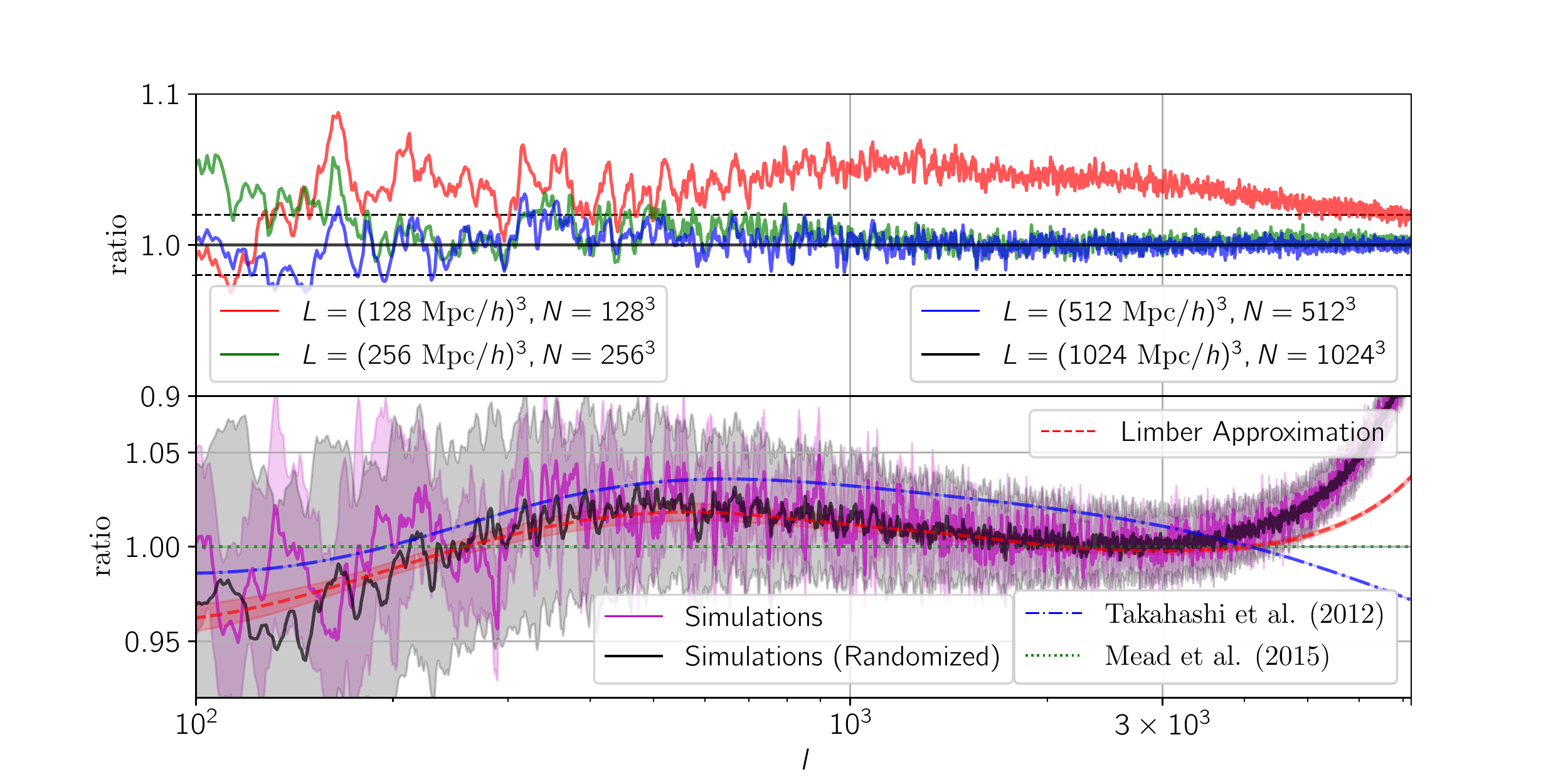}
	\caption{\emph{Top:} Comparison of the angular power spectra from different box-lengths ($L$) of the simulations, corresponding to $6^3$ (black), $12^3$ (blue), $24^3$ (green), and $48^3$ (red) replications for the light-cone construction. The blue case with $L=512$ Mpc/h and $12^3$ replica is used for the main analysis. \emph{Bottom}: Angular power spectrum of the randomised and non-randomised convergence maps relative to the results obtained from the three dimensional matter power spectra and the Limber approximation (all for the simulation with $L = 512$ Mpc/h). The results of \citet[][]{takahashi_2012} and \citet[][]{mead_2015} (also obtained via the Limber approximation) are shown for comparison.}
	\label{fig:cl_replication}
\end{figure}

The goal of this appendix is to show that the results or the paper are unaffected by the box-replication procedure that we apply to the $N$-body outputs in order to construct full-sky light-cones and convergence maps. We test the replication procedure by running simulations with four different box-sizes leading to different replication counts. The angular power spectra of the simulations with box-lengths $L=128,\, 256,\, 512,\, 1024$ Mpc/h with and without randomization are illustrated in Fig.~\ref{fig:cl_replication}.

The top panel of Fig.~\ref{fig:cl_replication} shows the angular power spectra measured from the convergence maps that were obtained via the light-cone construction discussed in Sec.~\ref{sec:convergence_maps}. Different lines refer to different box-sizes ($L$) and therefore different numbers of box replica (since all light-cones cover the same range in redshift of $z=0.1-1.5$). The plot shows that the setup used in the main analysis (i.e. the blue line with box-size $L = 512$ Mpc/h and $12^3$ replica) is well converged to the level of a few percent.

In the bottom panel of Fig.~\ref{fig:cl_replication} we compare the randomised to the non-randomised version of the light-cone production, again for the standard case of $L = 512$ Mpc/h and $12^3$ replica. The angular power spectra of both versions agree very well except for a small ($\sim2$ percent) difference at the larges physical scales. Furthermore, both power spectra are in good agreement with the one obtained via the Limber approximation and the three dimensional matter power spectra from the $N$-body outputs.

\begin{figure}[tbp]
	\centering
	\includegraphics[width=\textwidth,trim=1.1cm 0.18cm 0.2cm 0.65cm,clip]{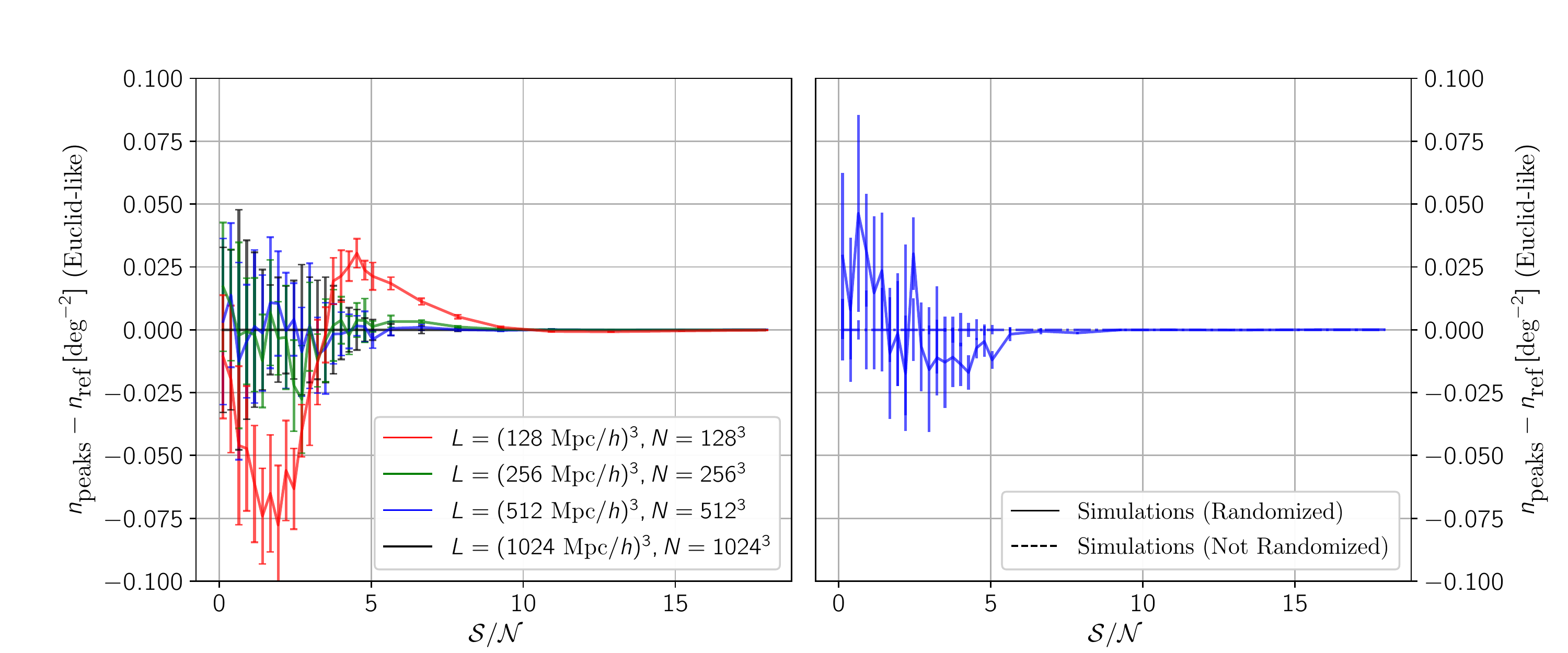}
	\caption{Absolute effects of the replication and randomization process on the peak distribution for the Euclid-like survey configuration and a $\theta_k=2$ arcmin smoothing scale. \emph{Left:} Peak differences of all four different box sizes relative to the $L=1024$ Mpc/h case. \emph{Right}: Peak difference between the randomised and non-randomised convergence maps for the $L=512$ Mpc/h case.}
	\label{fig:peaks_replication}
\end{figure}

In a second step we perform the same type of test for the peaks of the convergence maps, using the Euclid-like survey configuration and a $\theta_k=$ 2 arcmin smoothing scale. Fig.~\ref{fig:peaks_replication} shows the resulting peak differences for the different box-sizes (left) and the randomised/non-randomised case (right). Comparing this result to Fig. \ref{fig:peaks_dmo_diff} we see see the effect of the replication or randomization on the peak counts is negligible compared to the cosmological signal.

From testing the angular power spectrum and the peak statistics we conclude that errors due to box-size, numbers of replications, or the randomization method remain subdominant over all relevant ranges of scales. Furthermore, since we mainly focus on the relative effects of the baryonic versus dark-matter-only cases, all potential sources of errors investigated above are expected to cancel out.

\section{Resolution Effects}
\label{sec:mass_resolution}

\begin{figure}[tbp]
	\centering
	\includegraphics[width=\textwidth,trim=1.5cm 0.18cm 1.3cm 0.65cm,clip]{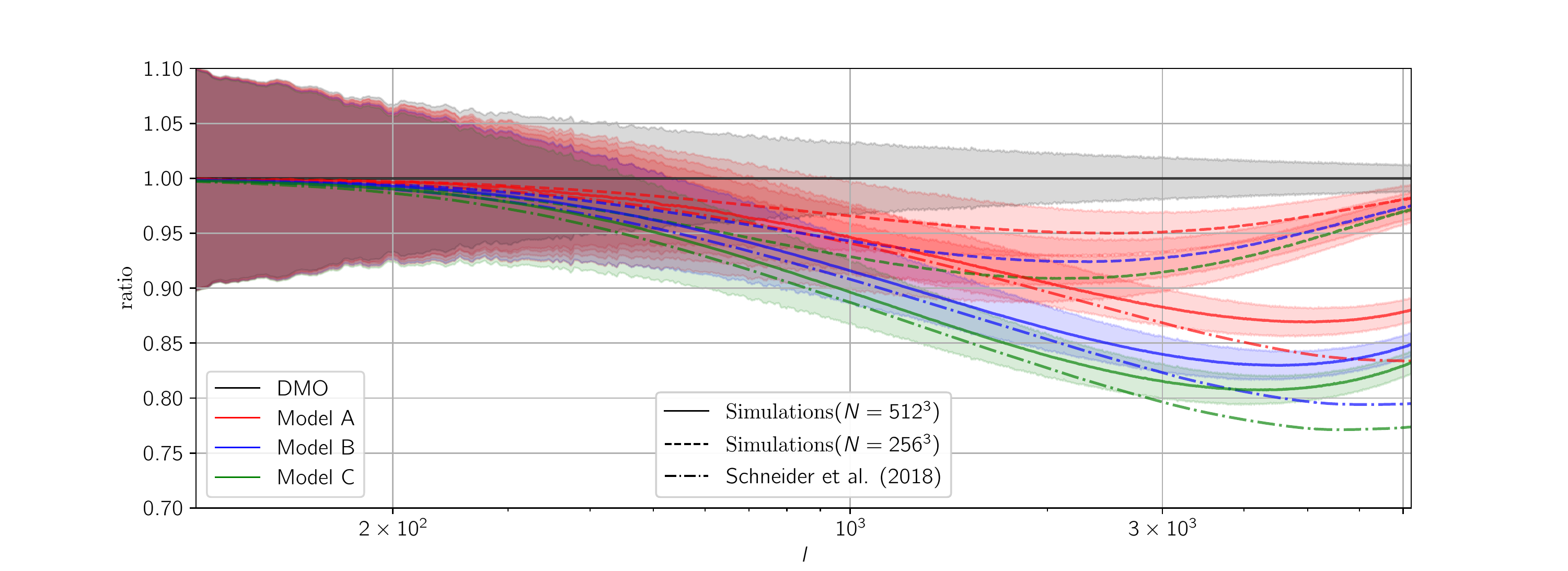}
	\caption{Relative effect of baryons on the angular power spectrum of the weak-lensing convergence. We show our simulations (solid), a lower resolution simulation (dashed) and the results from ST19 (dashed and dotted). Otherwise the Figure is the same as Fig.~\ref{fig:cl_baryons}.}
	\label{fig:cl_baryons_n256}
\end{figure}

As mentioned in Section \ref{sec:n_body_simulations}, we only modify haloes with more than 100 particles, i.e. with a halo mass above $M_\text{halo} = 9 \times 10^{12}$ M$_\odot$/h. This is not enough to get fully converged power spectra up to the scales relevant for future weak-lensing surveys \citep[see S15][]{schneider_2015}. In Fig.~\ref{fig:cl_baryons_n256} we show that the angular power spectrum from our simulations (solid lines) are not identical to the converged results from ST19. There is a difference of a few percent above $l=3000$ which can be attributed to the fact that baryon effects in haloes below our resolution limit contribute to the matter power spectrum at small scales. Note that the percent level difference at lower at $l<3000$ is not a resolution effect but is a result of the fact that the dash-dotted lines were derived using the Limber approximation (see S19). For comparison, we also plot the case of a lower mass resolution (of $N=256$ with the same box-size $L=512$ Mpc/h), which shows a significantly weaker baryon suppression (dashed lines). Although our results are not converged at the percent level for $l>3000$, we estimate the results to be good enough for the purpose of this paper.

\begin{figure}[tbp]
	\centering
	\includegraphics[width=\textwidth,trim=1.5cm 0.15cm 1.1cm 0.65cm,clip]{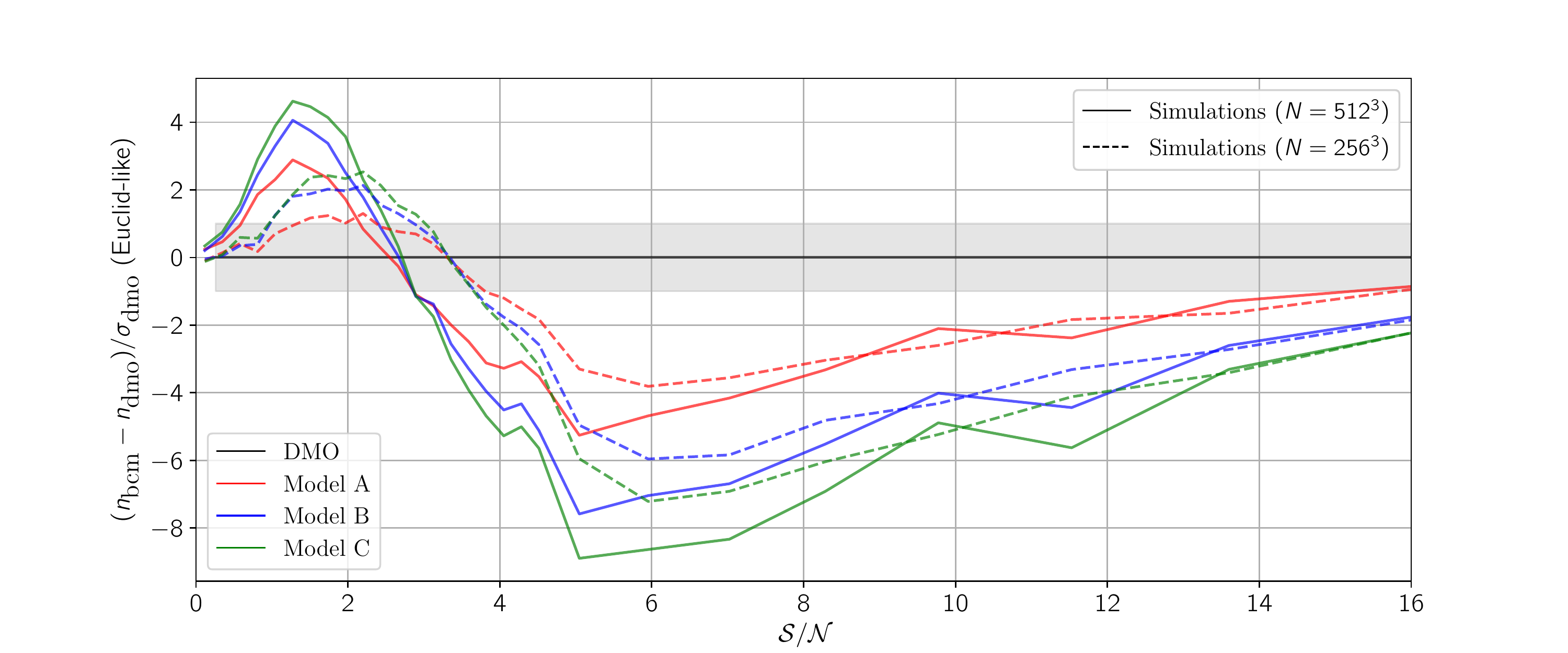}
	\caption{Significance of baryonic effects on the peak numbers of weak-lensing convergence maps
assuming an Euclid-like survey configuration, one single smoothing scale and two different mass resolutions. This result shows that significance is converged for the highest peaks, while it is not for the lower ones. }
	\label{fig:peaks_baryons_n256}
\end{figure}

Finally we investigate potential resolution effects on the peak statistics. Fig.~\ref{fig:peaks_baryons_n256} shows the absolute and relative effect of all three benchmark models with both simulations using the Euclid-like survey and a $\theta_k=2$ arcmin smoothing scale. The solid lines correspond to the standard resolution ($N=512^3$) used in the main analysis, while the dashed lines show results of a reduced resolution ($N=256^3$). The figure shows that while the high peaks are fully converged this is not necessarily the case for the lower peaks. Unfortunately, we do not have an even higher resolution to establish if there is a remaining discrepancy. 

\end{document}